\documentclass[aip,reprint]{revtex4-1}

\usepackage{datetime}
\usepackage{graphicx}

\begin{document}

\title{Dielectric relaxation and crystallization behaviour of amorphous nilutamide}

\author{N. S. K. Kumar}
\affiliation{Department of Physics, School of Physical, Chemical and Applied Sciences, Pondicherry University, Puducherry-605014, India}

\author{G. Govindaraj}
\email[Corresponding author, Email:]{ ggraj\_7@yahoo.com}
\affiliation{Department of Physics, School of Physical, Chemical and Applied Sciences, Pondicherry University, Puducherry-605014, India}

\author{U. Sailaja}
\affiliation{Department of Physics, M.E.S. Keveeyam College, Valancherry, Malappuram, Kerala-676552, India}

%\date{\today\hspace{5mm}\currenttime}
\date{\today}

\begin{abstract}
The molecular mobility of glass and supercooled liquid states of nilutamide has been studied with broadband dielectric spectroscopy for a wide range of temperature and frequency. Besides primary $\alpha$-relaxation an excess wing like secondary relaxation is observed. The temperature dependence of structural $\alpha$-relaxation show non-Arrhenius behaviour, and follows Vogel-Fulcher-Tammann (VFT) empirical formula. The glass transition temperature, T$_{g}$=302K and fragility index, m=76 are obtained from the VFT parameters. The structural $\alpha$-relaxation process is non-Debye with Kohlraush-Williams-Watts stretched exponential $\beta_{KWW}$=0.76.  Secondary relaxation time of nilutamide coincides with the primitive relaxation time calculated from the coupling model. Hence the secondary relaxation process of nilutamide is treated as the Johari-Goldstein (JG) $\beta$-process, which is the precursor of the structural $\alpha$-process.  Recent report on nilutamide indicates the increase of nucleation even below T$_{g}$, however we attributed to the JG $\beta$-relaxation. During the dielectric measurements amorphous nilutamide recrystallizes. The crystallization of amorphous nilutamide has been studied by the isotherm dielectric measurements at T=326K over a period of time. The crystallization follows Avrami equation and the parameters are obtained.
\end{abstract}


\maketitle %\maketitle must follow title, authors, abstract and \pacs

\section{Introduction}
Pharmaceuticals in amorphous form are more relevant due to their improved solubility and bioavailability as compared with the crystalline form, which are poorly water soluble. About 40\% of pharmaceuticals are poorly water soluble and results in inadequate bioavailability\cite{AmorpPharm-2002}. Amorphous materials are metastable with tendency to recrystallize. It was noted that even the sample kept many degrees below the glass transition temperature, may undergo recrystallization during the storage time\cite{AlieMolMobPharm-2004,StabilityPharm-1998}. A key factor which affect the stability of amorphous phase and tendency towards crystallization is molecular mobility\cite{MolMobPharm-2010,MolMobPharm-2012,MolMobPharm-2002}.
\par
Broadband dielectric spectroscopy is widely used technique to study dielectric relaxation and molecular dynamics of glasses, supercooled liquids, polymers and even disordered semiconductors\cite{BDS2002,KKPhysicaB}. The dielectric relaxation studies of pharmaceutical samples with broad band dielectric spectroscopy is very useful since the knowledge of molecular dynamics may leads to improve bioavailability and stability\cite{Craig1995,SailajaKetoprofen,SailajaFenofibrate}. The molecular motions responsible for the crystallization from the glass are an important topic. Hence, the study of molecular dynamics through dielectric relaxation of liquid and glassy phase of pharmaceuticals are very common\cite{DrugDelRev2015,SailajaKetoprofen,SailajaFenofibrate}. The existed crystallization model proposes that the molecular process responsible for crystal growth in glassy state is from the local molecular motions\cite{CrystallizationModel1,CrystallizationModel2}. Crystal nucleation requires motion of the entire molecule and Johari Goldstein (JG) $\beta$-relaxation is credited to the initial process of crystallization\cite{Ngai_Book01}.
\par
Nilutamide is one of the relevant active pharmaceutical drugs used for the treatment of prostate cancer. Nilutamide is one of the compounds, which is poorly water soluble in crystalline form. Nilutamide is classified as slowly crystallizing compound\cite{TaylorClassifica_01,TaylorClassifica_02}. Nucleation and crystal growth of amorphous nilutamide has been investigated using DSC, FTIR, Raman spectroscopy, NMR by Transi and Taylor\cite{Nilu_CrystEnggComm2014} and reported unusual case of increased nucleation rate of metastable poly morph even at temperature below T$_{g}$. However, no investigation is reported in literature for the molecular motion of liquid and glassy states of nilutamide.  In the present work, dielectric relaxation of supercooled liquid and glassy states of nilutamide is studied with broadband dielectric spectroscopy. The crystallization of amorphous nilutamide studied with the isotherm dielectric measurements at temperature 326 K. This is the first attempt to describe the molecular mechanism and crystallization of amorphous nilutamide using broadband dielectric spectroscopy.

\section{Materials and Methods}

\subsection{Materials}
\begin{figure}[h!]
\includegraphics[width=70mm]{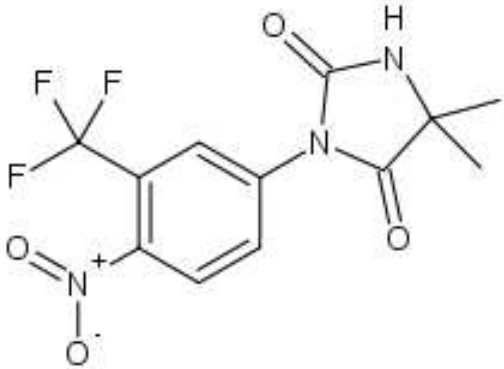}
\caption{Molecular structure of nilutamide}
\label{fig:struct}
\end{figure}

Nilutamide (5,5-dimethyl-3-[4-nitro-3-(trifluoromethyl)phenyl]imidazolidine-2,4-dione) with chemical formula C$_{12}$H$_{10}$F$_{3}$N$_{3}$O$_{4}$, in crystalline powder form was purchased from Sigma Aldrich and used for the experimental studies. The molecular structure of nilutamide is shown in Fig. \ref{fig:struct}.

\subsection{Methods}

\subsubsection{Raman spectroscopic measurements}
Raman spectroscopic measurement of crystalline and amorphous nilutamide has been obtained with 514 nm radiation. Raman shift of the range of 100 to 1800 cm$^{-1}$ has been obtained. Raman spectrum of crystalline (powder) sample is obtained at room temperature. Crystalline powder of the sample is melted by heating above melting temperature ($T_{m}$=426 K), and then it is quenched to 153 K with cooling rate of 15 K/min. Hence formed nilutamide glass is subjected to Raman measurements. Raman measurements of glassy and supercooled nilutamide have been carried out for various temperatures from 153 K to 348 K.  
\subsubsection{Broadband dielectric spectroscopy measurements}
The dielectric measurement was performed using stainless steel sample cell with 0.2 mm thick and 30mm diameter. Teflon was used as the spacer for the electrodes. Complex dielectric data was taken for a frequency range of $10^{-2}$Hz to $10^{7}$Hz for various temperature with alpha analyser of Novocontrol broad band dielectric spectrometer which has a temperature controller of nitrogen gas cryostat with a temperature stability of 0.1 K. The sample was carefully filled in the sample cells and heated to slightly above the melting temperature 383 K for 20 minutes and made into melt. It is pressed well and avoids formation of bubbles. It is subjected to fast cooling with a cooling rate of 15 K/min to a temperature of 123 K. Complex dielectric measurements of glassy and supercooled sample have been carried out for a temperature range 193 K to 338 K in proper steps.

\section{Results and discussion}

\subsection{Raman spectra}
\begin{figure}[h!]
\includegraphics[width=80mm]{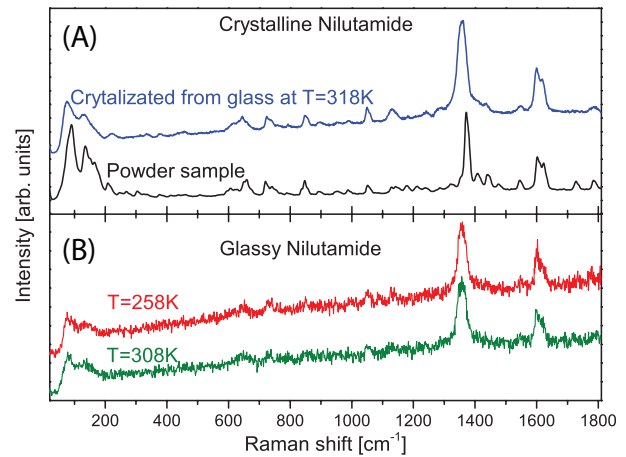}
\caption{(A). Raman spectra of nilutamide in crystalline (powder) and recrystallized form (B). Raman spectra of glassy form of nilutamide at two typical temperatures.}
\label{fig:Raman}
\end{figure}

The crystalline powder form of nilutamide shows all characteristic peaks of stable nilutamide as shown in Fig. \ref{fig:Raman}(A). Nilutamide is observed in stable and metastable forms. Melt quenched nilutamide show broader peaks and do not show significant variation for the temperature range of 153 K to 308 K. Raman spectra of glassy nilutamide for typical temperatures are shown in Fig. \ref{fig:Raman}(B). Raman spectra at temperature 318 K and above show show well defined peaks as shown in Fig. \ref{fig:Raman}(A). The glassy nilutamide recrystallized in the metastable form. For the stable form the most intense peak occurred at 1372 cm$^{-1}$ while for metastable form it occurred at 1357 cm$^{-1}$ very much similar as observed by Transi and Taylor\cite{Nilu_CrystEnggComm2014}. Initial powder form show stable nilutamide while glassy and recrystallized materials show metastable form. Two peaks at 1602 cm$^{-1}$ and 1622 cm$^{-1}$ in the stable form are shifted to lower wavenumbers in the metastable forms (1598 cm$^{-1}$ and 1616 cm$^{-1}$ respectively). The metastable also show a notable decrease in the intensity of carbonyl stretching bands between 1700 - 1800 cm$^{-1}$ with 1700 cm$^{-1}$ peak being much weaker as in the earlier studies\cite{Nilu_CrystEnggComm2014}. Raman spectroscopy of glassy and crystalline states of nilutamide studies show the amorphous nilutamide is more similar to the meta-stable states of nilutamide and it recrystallizes to the meta-stable states above T=318 K.

\subsection{Dielectric relaxation studies}
\begin{figure*}[th!]
\includegraphics[width=170mm]{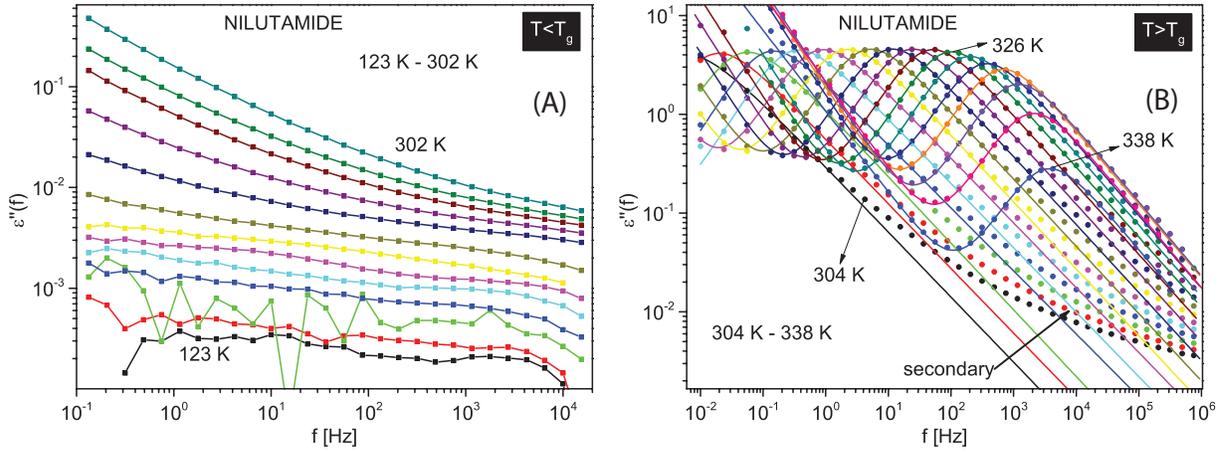}
\caption{(A). The dielectric loss data below T$_{g}$ do not show well defined secondary processes. (B). The dielectric loss data above T$_{g}$ structural $\alpha$ relaxation process, excess wing like secondary process and dc conductivity due to translational motion of ions. The dielectric loss data are fitted with $\sigma_{dc}$ and HN function for the $\alpha$ relaxation process.}
\label{fig:eps_data}
\end{figure*}

\begin{figure}[th!]
\includegraphics[width=80mm]{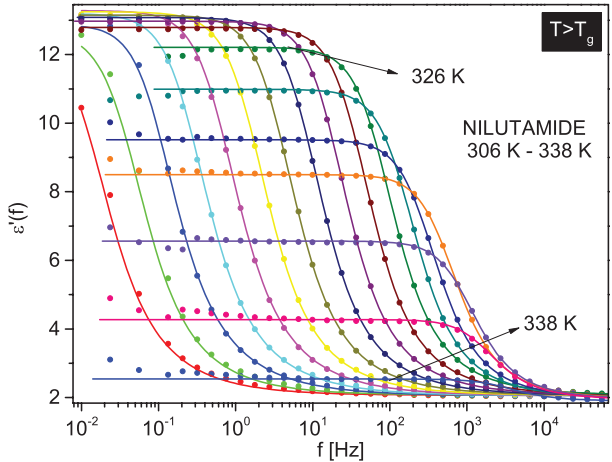}
\caption{Real part of dielectric permittivity at temperature above T$_{g}$. Nilutamide show a dramatic change in dielectric strength above the temperature 326 K, as an indication of crystallization of the sample.}
\label{fig:eps_data_re}
\end{figure}

The dielectric loss data of glassy and supercooled nilutamide, below and above T$_{g}$ are shown in Fig. \ref{fig:eps_data}(A) and (B) respectively.
The real part of complex dielectric data is shown in Fig. \ref{fig:eps_data_re}.  
The dielectric loss data below T$_{g}$ does not show well resolved secondary relaxation process. The dielectric loss data above T$_{g}$ show the structural $\alpha$-relaxation process and excess wing like secondary process. Dielectric loss at high temperatures also constitutes the dc conductivity contribution due to the translational motion of free ions. Generally the complex dielectric permittivity of can be fitted with,
\begin{eqnarray}
\varepsilon^{*}(\omega)&&=-i\frac{\sigma_{dc}}{\varepsilon_{0}\omega}+\frac{\Delta\varepsilon_{\alpha}}{(1+(i\omega \tau_{HN})^{\alpha_{HN}})^{\beta_{HN}}}    \nonumber  \\
&&+\frac{\Delta\varepsilon_{\beta}}{(1+(i\omega \tau_{CC})^{(1-\alpha_{CC})})}+\varepsilon_{\infty}
\label{eqn:eps_fit}
\end{eqnarray}
where $\sigma_{dc}$ is the dc conductivity, which contributes $\sigma_{dc}/(\varepsilon_{0}\omega)$ to the dielectric loss spectra. $\varepsilon_{\infty}$ is the high frequency limit of dielectric constant. Primary $\alpha$-process originates due the structural relaxation with cooperative motions of molecules are modelled with the asymmetric function Harviliak-Negami\cite{HNoriginal1967} (HN), with dielectric strength $\Delta\varepsilon_{\alpha}$, relaxation time $\tau_{HN}$ and shape parameters $\alpha_{HN}$ and $\beta_{HN}$. The secondary process are generally modelled with symmetric function Cole-Cole, with dielectric strength $\Delta\varepsilon_{\beta}$, relaxation time $\tau_{CC}$ and shape parameter $\alpha_{cc}$. The dielectric relaxation time of $\alpha$ process $\tau_{\alpha}$ is obtained as $\tau_{max}$, calculated from fit parameters of HN function\cite{RichertHNPeak1994}. Temperature dependence of relaxation time and viscosity of the supercooled liquids generally show non-Arrhenius behaviour.
\begin{figure}[h]
\centering
\includegraphics[width=60mm]{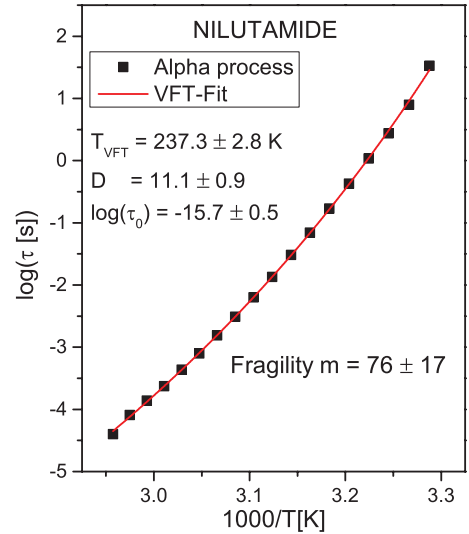}
\caption{The structural relaxation, $\alpha$-process of nilutamide shows non-Arrhenius behaviour and follows Vogel-Fulcher-Tamman (VFT) equation.}
\label{fig:vft}
\end{figure}
The structural relaxation, $\alpha$-process is thermally activated as shown in Fig. \ref{fig:vft}, which follows empirical Vogel-Fulcher-Tammann (VFT) equation\cite{VFT1-1921,VFT2-1923,VFT3-1926}.
\begin{equation}
\tau_{\alpha}=\tau_{\infty} exp[DT_{0}/(T-T_{0})]
\label{eqn:VFT}
\end{equation}
where D is the strength parameter, T is absolute temperature, $\tau_{0}$ is the high temperature limit of $\tau_{\alpha}$ and T$_{0}$ is the VFT temperature at which all important molecular motion would cease and relaxation time becomes infinity. The glass transition temperature is a kinetic transition that occurs when the system falls out of equilibrium due to the increasingly restricted molecular mobility and increasing viscosity, resulting in a glassy state\cite{GlassTransiPharmaReview,GlassTransiDSC}. The temperature at which relaxation time $\tau_{\alpha}$=100s are generally considered as glass transition temperature\cite{DrugDelRev2015}. T$_{g}$ of nilutamide is obtained as 302$\pm$11 K from the VFT fit parameters and it is close agreement with the $T_{g}$ obtained from DSC measurement by Transi and Taylor\cite{Nilu_CrystEnggComm2014}, which is 305 K. Materials showing Arrhenius behaviour are termed as ``strong" and which exhibits non-Arrhenius dependency are referred ``fragile". Fragility index is generally obtained from the Angell plot, which quantifies the amount of deviation of relaxation time from Arrhenius behaviour\cite{Angell_fragility}. The fragility index m, can be obtained from the VFT parameters as \cite{fragility_VFT_Eqn}
\begin{equation}
m=\left.\frac{dlog_{10}\tau_{\alpha}}{d(T_{g}/T)}\right|_{T=T_{g}}=D\frac{T_{0}}{T_{g}}\left(1-\frac{T_{0}}{T_{g}}\right)^{-2}log_{10}e
\label{eqn:Fragility}
\end{equation}
The typical value of fragility is between 16 and 200. It is considered that the fragility index plays an important role to establish an appropriate storage condition for the particular drug since it correlate with the glass-forming ability and physical stability of amorphous systems. Fragility index describe the rapidity with which a liquid structure, which was arrested at the glass transition during cooling, becomes disrupted on heating\cite{DrugDelRev2015}. The fragility index of nilutamide, m=76$\pm$17 is obtained from Eq. \ref{eqn:Fragility} and VFT parameters. The error in fragility is obtained from the error in VFT parameters. Nilutamide can be classified as the intermediate glass-former drug. The ratio $T_{m}/T_{g}$=1.41, and the difference, $T_{g}-T_{0}$=64.7, are also indications of the system is fragile. The fragility of nilutamide is comparable with other glass forming pharmaceuticals like, glibenclamide\cite{Glibenclamide} (m=78), indapamide\cite{Indapamide} (m=76). It is worth noting that many other molecular glass formers like sqalane\cite{Sqalane} (m=75), phenolphthalein dimethylether\cite{PDE} (m=75), isoeugenol\cite{Isoeugenol} (m=73) has the similar fragility. 
\par
\begin{figure}[h!]
\centering
\includegraphics[width=80mm]{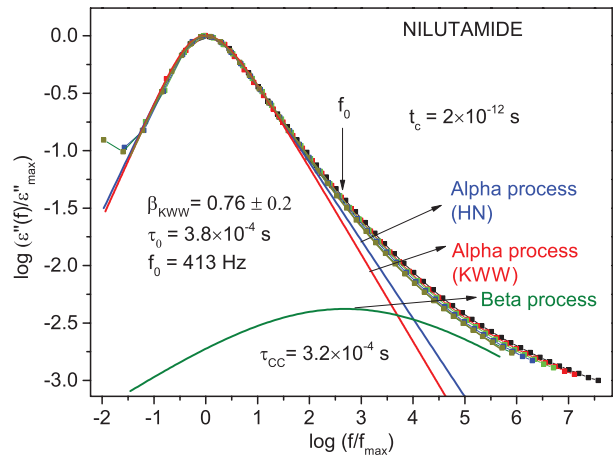}
\caption{The scaled dielectric loss data of nilutamide, near glass temperatures with KWW fit (red curve). The dielectric loss data are fitted with Harviliak-Negami function for $\alpha$-process (blue curve) and Cole-Cole function for secondary, $\beta$-process (green curve). The primitive relaxation time $\tau_{0}$ obtained from the Ngai's coupling model corresponds a frequency f$_{0}$ and the secondary process is almost near to the primitive relaxation indicating the secondary relaxation process is a Johari-Goldstein process.}
\label{fig:KWWCoupling}
\end{figure}

Debye dielectric relaxation function\cite{Debye-1912} is given by $\phi(t)=exp(-t/\tau_{D})$ where $\tau_{D}$ is the Debye relaxation time and it produces a polarization current density $j_{p}$ $\propto$ $-d\phi_{p}(t)/dt$. Corresponding to this relaxation function, frequency dependent complex permittivity is obtained as $\varepsilon^{*}(\omega)=\varepsilon_{p}/(1+i \omega \tau_{D})$. Molecular relaxation process in almost all glass-forming materials are non-exponential and it is modelled with the stretched exponential function known as Kohlarush-Williams-Watts (KWW) function\cite{Ngai_Book01}.
\begin{equation}
\phi(t)=exp\left[-\left(\frac{t}{\tau_{KWW}}\right)^{\beta_{KWW}}\right]
\label{eqn:KWW}
\end{equation}
The KWW lacks a analytical function in frequency domain for all values of exponent $\beta_{KWW}$ except for $\beta_{KWW}$=0.5. Taking one side Fourier transform of the KWW is common for analysing the dielectric data. The non-exponential behaviour of the dielectric relaxation is attributed to the distribution of molecular relaxation times. The scaled dielectric loss of near glass transition temperature show KWW exponent has value $\beta_{KWW}=0.76$ as shown in Fig.  \ref{fig:KWWCoupling}.
The value of $\beta_{KWW}$ is comparable with other pharmaceutical materials glibenclamide\cite{Glibenclamide} ($\beta_{KWW}$=0.74), indapamide\cite{Indapamide} ($\beta_{KWW}$=0.72), acetaminophen\cite{Acetaminophen} ($\beta_{KWW}$=0.79). Shamblin \textit{et al.}, suggested that the distribution of structural relaxation time may raise the crystal nucleation rate, and consequently the shelf life of the pharmaceutical\cite{Shamblin_KWW_Nucleation}. It has been reported that the both physical and chemical stability decreases with the decrease of value of $\beta_{KWW}$. The faster modes of molecular motions within the distribution of relaxation time can be responsible for the nucleation in the glassy state, therefore the glass formers with smaller $\beta_{KWW}$ would be more susceptible to nucleate\cite{Johari_KWW_Nucleation}. But it should be noted down many of the pharmaceutical materials do not show the expected correlation between $\beta_{KWW}$ and tendency of crystallization\cite{DrugDelRev2015}. The stretching parameter $\beta_{KWW}$ has been often considered as a measure of degree of cooperativity and length scale or dynamic heterogeneity of molecular mobility reflected in the structural relaxation\cite{DrugDelRev2015}. Thus $\beta_{KWW}$ is also refereed as alternative parameter to the fragility index m in search of correlation with tendency of crystallization.
\par
Understanding the physical origin of secondary relaxation processes in glass forming materials is really challenging. It is widely accepted that the secondary process may originate either from the motion of the entire molecule or intermolecular process. The secondary process of intramolecular origin may be called $\gamma$-process. It was observed that there is a possible secondary relaxation process called Johari-Goldstein (JG) process, which are precursors of the primary structural relaxation process\cite{JG,Ngai_Beta_classification}. The JG relaxation is generally referred as the $\beta$-process and it is considered to be universal process in glass forming materials. Ngai has used a coupling model (CM) to identify the nature of the secondary process\cite{Ngai_Coupling1,Ngai_Coupling2}. The CM predicts a correlation of relaxation time of JG process, $\tau_{\beta}$ with the relaxation time of $\alpha$-process $\tau_{\alpha}$. The primitive relaxation time can be obtained from the empirical relation,
\begin{equation}
\tau_{0}=(t_{c})^{n}(\tau_{\alpha})^{(1-n)}
\label{eqn:Ngaicoupling}
\end{equation}
where, t$_{c}$ is the time characterizing the crossover from independent to cooperative fluctuation. For small glass formers its value is nearly 2$\times$10$^{-12}$ s. The coupling parameter, n=1-$\beta_{KWW}$ can be obtained from the structural $\alpha$ relaxation process. According to CM the scaled dielectric data should have primitive relaxation time $\tau_{0}=3.8\times$10$^{-4}$ s, which corresponds to the frequency of 413 Hz approximately. Generally secondary relaxation follows Arrhenius temperature dependence. To resolve secondary relaxation from the nearby $\alpha$ process is rather difficult, except some temperatures nearby T$_{g}$. Hence attempts to resolve $\alpha$-process and $\beta$-process, have made only for the scaled dielectric loss data at temperatures near T$_{g}$ as shown in Fig. \ref{fig:KWWCoupling}.  The master curve of dielectric loss is fitted using the Eq.\ref{eqn:eps_fit}, without dc conductivity contribution. The secondary relaxation follows Cole-Cole dielectric relaxation function with relaxation time $\tau_{CC}$=3.2$\times$10$^{-4}$ s, which is close agreement with the primitive relaxation time $\tau_{0}$ predicted by the coupling model. Hence the secondary relaxation observed in nilutamide can considered as the JG $\beta$-process ($\tau_{0}\approx\tau_{\beta}$), originates due to the motion of the entire molecule. Even though there is a possibility of hydrogen bonding in nilutamide the secondary relaxation process of the glass former is JG $\beta$-relaxation or closely resembling JG relaxation since it follows coupling model criteria\cite{Ngai_Beta_classification}.
\par
It is observed that the fragility factor m and stretched exponential factor $\beta_{KWW}$ are similar for glibenclamide\cite{Glibenclamide}, indapamide\cite{Indapamide} and present sample of interest nilutamide but with different VFT parameters and T$_{g}$. Nilutamide did not show explicit $\gamma$-process as in glibenclamide and indapamide. No $\gamma$-process is observed and it may be due to the absence of strong intramolecular motions of subgroups of nilutamide.
\par 
Paluch \textit{et al.}, found that the width of $\alpha$-loss peak near T$_{g}$ is strongly anticorrelates with the polarity of the molecule especially of van der Waals glass formers. Hence larger the dielectric loss strength $\Delta\varepsilon_{\alpha}$ of glass former, lesser the value of $\beta_{KWW}$ of the $\alpha$-relaxation process\cite{Ngai_eps_KWW_PRL_2016}. This has been described as the contribution of dipole-dipole interaction potential, V$_{dd}\propto$r$^{-6}$ (r is the distance between the molecules), to the intermolecular potential which makes the resultant potential more harmonic. Nilutamide has $\Delta\varepsilon_{\alpha}$=10.8 and $\beta_{KWW}$=0.76, near T$_{g}$. Even though there is a possibility of hydrogen in nilutamide, it shows similar behaviour as isooctylcyanobiphenyl\cite{Ngai_eps_KWW_PRL_2016}.
\par
Crystallization from organic undercooled melts is an important topic of investigation. Combined process of nucleation and crystal growth results the crystallization from glasses/supercooled liquids. As the temperature increases, the rate of crystal growth increases due to the increase of molecular mobility. Nucleation is a prerequisite to the crystal growth. Hence storing the sample at temperature where the nucleation rate is low will help to prevent the crystallization from glasses. Many materials show recrystallization during the storage time, even though they kept several degrees below T$_{g}$\cite{AlieMolMobPharm-2004,StabilityPharm-1998}. The molecular motions responsible for the secondary relaxation may be the key factor for such tendency of crystallization\cite{MolMobPharm-2010,MolMobPharm-2012,MolMobPharm-2002}.
\par
\begin{figure}[th!]
\includegraphics[width=80mm]{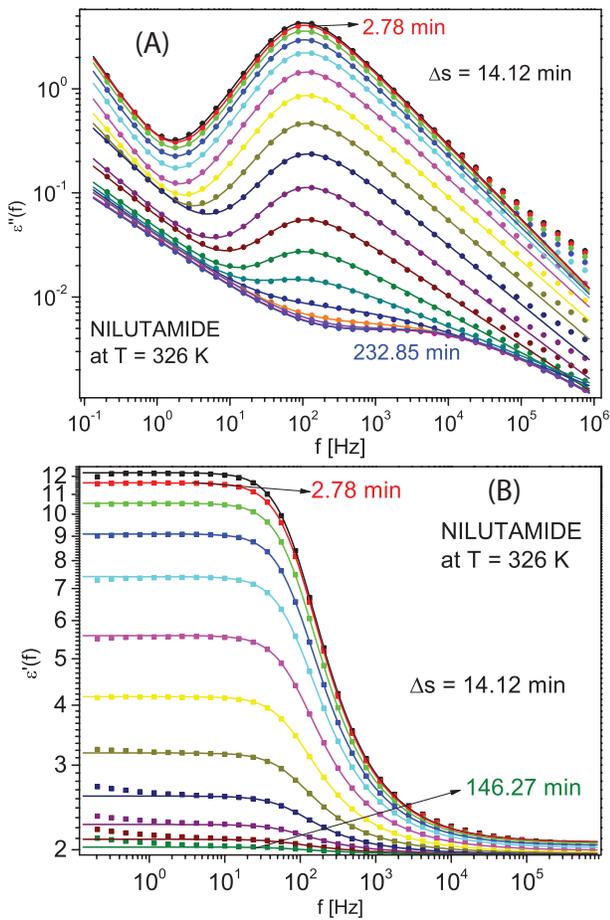}
\caption{(A). Imaginary and (B). real part of permittivity at temperature 326 K for various time. The dielectric data at initial, 2.78 min. and time interval of 14.12 min. are shown in the figure.}
\label{fig:eps_cryst}
\end{figure}

Nilutamide DSC measurements do not show recrystallization due to the rate of nucleation and/or crystal growth is slow relative to the time scale of the DSC measurements\cite{Nilu_CrystEnggComm2014}. It is observed that the melt had to be cooled to at least 243 K for crystallization to observed upon heating\cite{Nilu_CrystEnggComm2014}. Raman and dielectric measurements has been carried out to the quenched melt below 153 K. During dielectric spectroscopic measurements the amorphous nilutamide show recrystallization during heating above the temperature of 326 K. The dielectric strength of $\alpha$-process decreases with increase of temperature above 326 K as shown in Fig. \ref{fig:eps_data_re}. This indicates that the molecule have sufficient kinetic energy at this temperature to start the crystal growth. The dramatic change in the dielectric strength of $\alpha$-process is a signature of the crystallization phenomena. During crystallization the relaxation time $\tau_{\alpha}$ is almost same for considerable period of time. The crystalline region of the material do not contribute to the dielectric response of $\alpha$-process. The density of mobile dipoles decreases and hence the intensity of the structural relaxation. The dielectric measurements at constant temperature 326 K at ambient pressure show a decrease in dielectric strength with time as an indication of increase of crystal growth and it is shown in Fig. \ref{fig:eps_cryst}.
\par
\begin{figure}[h!]
\includegraphics[width=60mm]{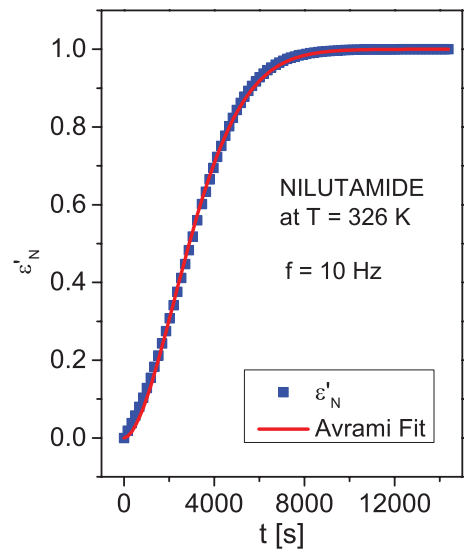}
\caption{Crystallization fraction as obtained from Eq.\ref{eqn:Avrami} for the dielectric data at 10 Hz, of the crystallizing nilutamide at temperature of 326 K.}
\label{fig:Avrami}
\end{figure}

The amount of crystallization can be estimated from the dielectric spectra using the normalized change in the dielectric dispersion $\varepsilon'_{N}$(t)
\begin{eqnarray}
&&\alpha(t)=\varepsilon'_{N}(t)=1-\frac{\Delta\varepsilon(t)}{\Delta\varepsilon(0)}=1-\frac{\varepsilon'(t)-\varepsilon'(\infty)}{\varepsilon'(0)-\varepsilon'(\infty)} \nonumber  \\
&&=1-exp[-(kt)^{n}] \label{eqn:Avrami}
\end{eqnarray}
where the conversion degree $\alpha$ is a measure of how rapidly active dipoles decreases during the crystallization process and it is connected to Avrami function\cite{DrugDelRev2015,API_Crystalization2015,Avrami}. The parameter k is the overall crystallization rate constant which includes both nucleation and growth. Avrami exponent n, correlates with the nature of crystal growth. The crystallinity fraction obtained from Eq. \ref{eqn:Avrami} for the dielectric data at 10 Hz of crystallizing nilutamide is shown in Fig. \ref{fig:Avrami}.
The Avrami exponent n has a value 1.76$\pm0.01$ at temperature 326K at ambient pressure. The crystallization rate constant k is obtained as 2.81$\pm0.01\times$10$^{-4}$ s$^{-1}$. Nilutamide has a fairly rapid growth rate. The good glass forming ability and lack of crystallization during cooling from melt may be from its low nucleation tendency. The Raman measurements show that the recrystallized sample is in the metastable form. The physical parameters of nilutamide obtained from the dielectric relaxation studies are listed in Table \ref{table:VFTfit}.
\begin{table*}[th!]
\caption{Physical parameter obtained for nilutamide using dielectric relaxation studies.}
\label{table:VFTfit}
\footnotesize
\setlength{\tabcolsep}{18pt}
\begin{tabular}{c c c c c c c}
\hline
$T_{0}$ [K]   & D                   & log($\tau_{0}$)  & T$_{g}$ [K]  & 
Fragility     & $\Delta\varepsilon_{\alpha}(T_{g})$ & $\beta_{KWW}$\\
\hline
237.3$\pm$2.8 & 11.1$\pm$0.9        & -15.7$\pm$0.5    & 302$\pm$11   & 
76$\pm$17     & 10.8$\pm$0.2        & 0.76$\pm$0.02\\
\hline
\end{tabular}
\end{table*}
\par
Nilutamide has three hydrogen bond acceptors and one donor group. Transi and Taylor considered the hydrogen bond between the amide group and carbonyl group which results the formation of chain of molecules\cite{Nilu_CrystEnggComm2014}.
More extensive hydrogen bonding is found in the metastable form as compared to the stable form. One of the possible origins of the metastable form is the conformational polymorphs due to different dihedral angle formed by the plane of two rings of nilutamide. Calculations of Transi and Taylor, show that minimum energy occurs when the rings are at angle 30$^{\circ}$ and conformational analysis show the stable polymorph nilutamide molecule is not planar but the phenyl ring rotated relative to the imidazole ring by about 54.4$^{\circ}$ leading to a higher energy conformer that may be stabilized by interaction in crystal\cite{Nilu_CrystEnggComm2014}. Recent studies show ibuprofen nucleate into metastable form (conformational poly-morph) above and below the glass transition temperature\cite{Ibuprofen_belowTg,Ibuprofen_aboveTg}. There are possibilities of maximum nucleation rate at temperature much below T$_{g}$ and the nucleation is controlled by JG $\beta$-process, rather than viscosity\cite{nucleation_beta}. 
\section{Conclusion}
Important pharmaceutical drug, nilutamide is melt-quenched and studied the molecular dynamics of glassy and supercooled states using broadband dielectric spectroscopy. The dielectric loss data show structural $\alpha$-relaxation process and excess wing like secondary relaxation process. The $\alpha$-process is a thermally activated non-Arrhenius process which follows VFT equation. Glass transition temperature and T$_{g}$=302 K is obtained from the dielectric spectra by the assumption of $\tau_{\alpha}$=100 s at T$_{g}$. The VFT parameter of $\alpha$-process indicates nilutamide is a moderate fragile glass-former with fragility index m=76. The $\alpha$-process is a non-Debye with KWW stretched exponential parameter $\beta$=0.76, near glass transition temperatures. Based on Ngai's coupling model the secondary process is a Johari-Goldstein process which may be the precursor of the structural $\alpha$-process. Nilutamide do not show secondary $\gamma$-process and it may be due to the absence of strong intramolecular motions of subgroups. The dielectric data of nilutamide show, the strong anticorrelation of width of $\alpha$ relaxation process with the dielectric loss strength predicted by Paluch and coworkers. Nilutmaide crystallizes during the dielectric measurements and the strength of $\alpha$-process decreases during crystallization as the density of mobile dipole decreases. Isotherm dielectric measurements at 326 K have been carried out for various times. Avrami parameters n and k is found to be 1.76 and k=2.81$\times$10$^{-4}$ s$^{-1}$ respectively at temperature T=326 K. Non-crystallization during cooling from the melt and  good glass forming ability of the nilutamide may be attributed to the low nucleation tendency. The former report on the increased rate of nucleation below T$_{g}$ may be attributed to the observed JG $\beta$ relaxation of glassy nilutamide. 
%\linebreak
\begin{acknowledgments}
Authors thank Central Instrumentation Facility, Pondicherry University for the dielectric spectroscopy and Raman spectroscopic measurements. Authors acknowledge CSIR project F.03(1279)/13/EMR-II for financial support and one of the authors (NSKK) acknowledges for senior research fellowship.
\end{acknowledgments}

%\nocite{*}
\section*{References}
\bibliography{nilu}
\end{document}